\documentclass{aa-package/aa}  

\usepackage{natbib}

\usepackage{txfonts}
\usepackage{color, colortbl}
\usepackage[dvipsnames,table]{xcolor}
\definecolor{pinegreen}{RGB}{1, 121, 111}
\usepackage[hyperindex,pdftex,colorlinks,citecolor=pinegreen]{hyperref}
\usepackage{fix2col}
\usepackage{booktabs}
\usepackage{graphicx}
\usepackage{xspace}
\usepackage[font=small,labelfont=small,labelfont=bf,compatibility=false]{caption}
\usepackage{subcaption}
\usepackage{lscape}
\usepackage{tikz,pgfplots,tikz-3dplot}
\usepackage{hvfloat}
\usetikzlibrary{plotmarks}
\usetikzlibrary{shapes.geometric}
\usetikzlibrary{shapes,arrows}
\usetikzlibrary[pgfplots.groupplots]

\usepackage{array}
\newcolumntype{L}[1]{>{\raggedright\let\newline\\\arraybackslash\hspace{0pt}}m{#1}}
\newcolumntype{C}[1]{>{\centering\let\newline\\\arraybackslash\hspace{0pt}}m{#1}}
\newcolumntype{R}[1]{>{\raggedleft\let\newline\\\arraybackslash\hspace{0pt}}m{#1}}

\newcommand{\degree}{\ensuremath{{}^{\circ}}\xspace}

\renewcommand\eqref[1]{\ifnum\ifhmode\spacefactor\else2000\fi>1000 Equation~(\ref{#1})\else Eq.~(\ref{#1})\fi}
\newcommand\eqreftwo[2]{\ifnum\ifhmode\spacefactor\else2000\fi>1000 Equations~(\ref{#1}) and (\ref{#2})\else Eqs.~(\ref{#1}) and (\ref{#2})\fi}

\newcommand\figref[1]{\ifnum\ifhmode\spacefactor\else2000\fi>1000 Figure~\ref{#1}\else Fig.~\ref{#1}\fi}
\newcommand\figreftwo[2]{\ifnum\ifhmode\spacefactor\else2000\fi>1000 Figures~\ref{#1} and \ref{#2}\else Figs.~\ref{#1} and \ref{#2}\fi}
\newcommand\figrefthree[3]{\ifnum\ifhmode\spacefactor\else2000\fi>1000 Figures~\ref{#1}, \ref{#2}, and \ref{#3}\else Figs.~\ref{#1}, \ref{#2}, and \ref{#3}\fi}

\newcommand\secref[1]{\ifnum\ifhmode\spacefactor\else2000\fi>1000 Section~\ref{#1}\else Sect.~\ref{#1}\fi}
\newcommand\secreftwo[2]{\ifnum\ifhmode\spacefactor\else2000\fi>1000 Sections~\ref{#1} and \ref{#2}\else Sects.~\ref{#1} and \ref{#2}\fi}

\newcommand\tabref[1]{\ifnum\ifhmode\spacefactor\else2000\fi>1000 Table~\ref{#1}\else Table~\ref{#1}\fi}

\def\addlegendimage{\csname pgfplots@addlegendimage\endcsname}

\begin{document}  

\title{Long-term stability of the HR~8799 planetary system without resonant lock}
\author{Ylva G\"{o}tberg
          \inst{1,}\inst{2}
          \and
          Melvyn B.\ Davies\inst{1}
          \and 
          Alexander J.\ Mustill\inst{1}
          \and
          Anders Johansen\inst{1}
          \and
          Ross P.\ Church\inst{1}
          }
\institute{Lund Observatory, Department of Astronomy and Theoretical Physics, Lund University, Box 43, 22100 Lund, Sweden
         \and
         Anton Pannekoek Institute for Astronomy, University of Amsterdam, Science Park 904, 1098 XH, Amsterdam, The Netherlands 
             \email{Y.L.L.Gotberg{@}uva.nl}
             }
\date{Received ..... ; accepted ..... }

\abstract{
HR~8799 is a star accompanied by four massive planets on wide orbits. The observed planetary configuration has been shown to be unstable on a timescale much shorter than the estimated age of the system ($\sim 30$~Myr) unless the planets are locked into mean motion resonances. This condition is characterised by small-amplitude libration of one or more resonant angles that stabilise the system by preventing close encounters.
We simulate planetary systems similar to the HR~8799 planetary system, exploring the parameter space in separation between the orbits, planetary masses and distance from the Sun to the star. We find systems that look like HR~8799 and remain stable for longer than the estimated age of HR~8799. None of our systems are forced into resonances.
We find, with nominal masses ($M_{\rm{b}} = 5\, M_{\rm Jup}$ and $M_{\rm{c,d,e}} = 7\, M_{\rm Jup}$) and in a narrow range of orbit separations, that 5 of 100 systems match the observations and lifetime. Considering a broad range of orbit separations, we find 12 of 900 similar systems. The systems survive significantly longer because of their slightly increased initial orbit separations compared to assuming circular orbits from the observed positions. A small increase in separation leads to a significant increase in survival time. The low eccentricity the orbits develop from gravitational interaction is enough for the planets to match the observations. With lower masses, but still comfortably within the estimated planet mass uncertainty, we find 18 of 100 matching and long-lived systems in a narrow orbital separation range. In the broad separation range, we find 82 of 900 matching systems. 
Our results imply that the planets in the HR~8799 system do not have to be in strong mean motion resonances. 
We also investigate the future of wide-orbit planetary systems using our HR~8799 analogues. We find that 80\% of the systems have two planets left after strong planet-planet scattering and these are on eccentric orbits with semi-major axes of $a_1 \sim 10$~AU and $a_2 \sim 30 - 1000$~AU.  We speculate that other wide-orbit planetary systems, such as AB~Pic and HD~106906, are the remnants of HR~8799 analogues that underwent close encounters and dynamical instability.}

\maketitle

\section{Introduction}

\begin{figure}
\centering
\includegraphics[scale=1]{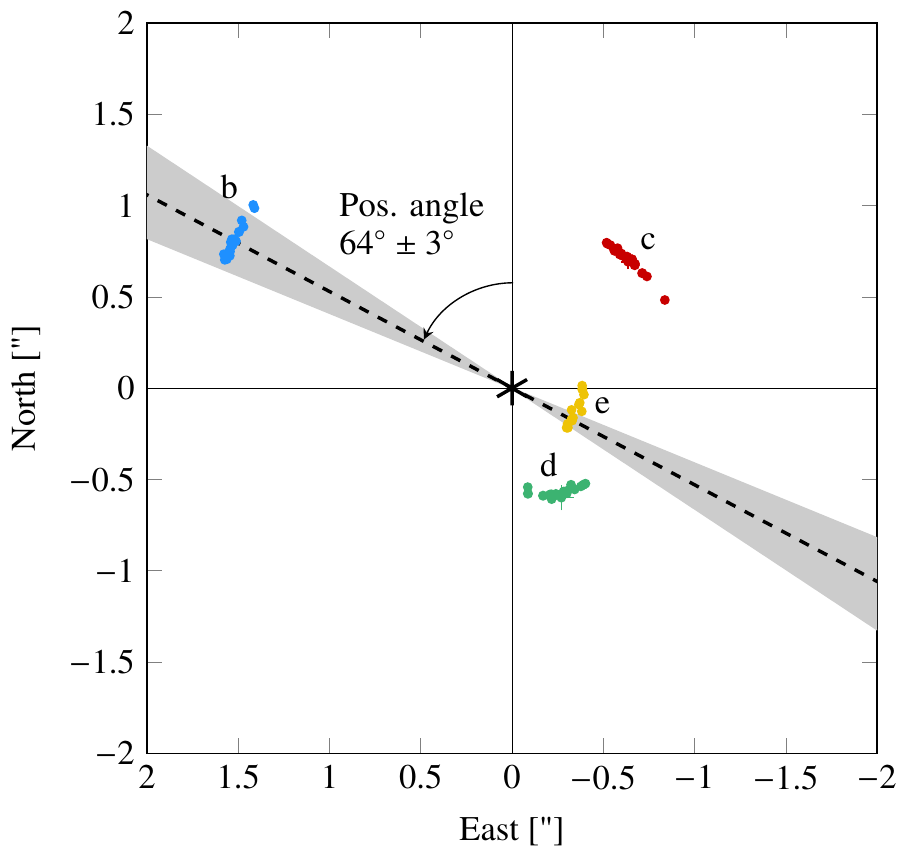}
\caption{Observations of the HR~8799 planets. The black dashed line corresponds to the position angle of the inclination as estimated by \cite{2014ApJ...780...97M} with the $\pm 1\sigma$ uncertainty shown by the grey region.}
\label{fig:obs_8799}
\end{figure}

HR~8799 (also called HD~218396 and HIP~114189) is a nearby \citep[$39.4 \pm 1.1$ pc;][]{2007A&A...474..653V}, young \citep[$30^{+20}_{-10}$~Myr assuming member of the Columba association;][]{2010Natur.468.1080M} A5V (up to F0V) star with mass $1.5 \pm 0.3\; M_{\odot}$ \citep{1999AJ....118.2993G, 2003AJ....126.2048G}. HR~8799 has four confirmed, directly imaged planets \citep{2008Sci...322.1348M, 2010Natur.468.1080M} and a debris disk \citep{2009ApJ...705..314S,2014ApJ...780...97M}. The planet masses are estimated to be $5$~(b) and $7$~(cde)~$M_{\text{Jup}}$ \citep{2010Natur.468.1080M}. The on-sky separations between planets and star are estimated to be 67.9~(b), 38.0~(c), 24.5~(d), and 14.5~(e) AU. \cite{2014ApJ...780...97M} estimated the inclination of the system with respect to the plane of the sky to be $26 \pm 3\degree$ by observing the outer part of the debris disk and assuming coplanarity with the planetary orbits. They estimate the position angle of the inclination to be $64 \pm 3\degree$ (G.~Kennedy, private communication). \figref{fig:obs_8799} shows all observations of the planets published at the time of writing. 

The estimates of the planetary masses in the HR~8799 system depend on the age of the system through cooling models, indicating higher planet masses with higher age \citep{2003A&A...402..701B}. The age of the HR~8799 system is uncertain and has been estimated using different techniques to range between 30~Myr and 1~Gyr \citep[see][and references therein]{2010Natur.468.1080M,2008Sci...322.1348M}. The planet mass estimates reach the brown dwarf regime when an age of $\gtrsim 100$~Myr is assumed. In this paper we consider HR~8799~bcde as planets; the case of brown dwarfs is more closely investigated by \cite{2010ApJ...721L.199M}. We assume the age 30~Myr in this paper as it is estimated from the Columba association. An age of 30~Myr leads to multi-Jovian masses that, however, are still in the planetary regime. Whether it is true that HR~8799 is associated with the Columba association could be revealed by Gaia in the near future \citep{2001A&A...369..339P}. Other uncertainties in system properties include distance \citep[$39.4\pm1.1$\,pc][]{2007A&A...474..653V} and stellar mass \citep[$1.5\pm0.3\mathrm{\,M}_\odot$][]{1999AJ....118.2993G,2003AJ....126.2048G}.

The formation of the massive HR~8799~bcde planets can be explained by the recently developed models of pebble accretion \citep{2012A&A...544A..32L}, but the dynamics of the system has been a puzzle unless resonances are considered. The orbits of the HR~8799~bcde planets are difficult to constrain as the planets have been observed during a small fraction of their orbits (see \figref{fig:obs_8799}). In recent work by \cite{2015ApJ...803...31P}, however,  fits are made to constrain the orbital elements and a variety of solutions are found, favouring a slightly more eccentric orbit for planet d. It is reasonable to believe that the planet orbits are close to circular, as with high eccentricity the orbits are likely to cross. However, the HR~8799 planetary system appears to be unstable on a timescale much shorter than the estimated age of the system when integrating the orbits of the planets, assumed initially circular, seen pole-on and non-resonant \citep{2010ApJ...710.1408F, 2009A&A...503..247R, 2012ApJ...755...38S}. A possible and popular explanation to the survival of the HR~8799 planetary system is that the planets are locked deep in strong mean motion resonances that can lead to stability timescales up to 1~Gyr \citep[e.g. ][]{2010ApJ...710.1408F, 2009MNRAS.397L..16G, 2009A&A...503..247R, 2010ApJ...721L.199M, 2011ApJ...729..128C, 2011ApJ...741...55S, 2013A&A...549A..52E,  2014MNRAS.440.3140G}. These resonant configurations are characterised by small-amplitude librations of one or more resonant arguments corresponding to two-, three-, or four-planet mean motion resonances, thus ensuring that systems remain protected for extended periods from close encounters between the planets. \cite{2013MNRAS.430..320M} find stabilising effects from the outer debris disk, although this requires a very massive debris disk.

In this paper we show simulations of planetary systems that look like the HR~8799 system, are stable for longer than the estimated age, and are not stabilised by strong resonant lock. We create these systems by starting the planetary orbits slightly more widely separated than the observed on-sky separation assuming circular orbits. A more widely separated configuration can also look like the HR~8799 system as the orbits are not exactly circular. Larger separation between the planetary orbits results in a much longer stability timescale, which makes systems remain stable for longer than the estimated age of HR~8799. This solution is different from previously published explanations, as we do not need protective resonant lock or any other stabilising effects to create a long-lived system.

In \secref{sec:simulations} we describe how we simulate planetary systems. \secreftwo{sec:HR8799unstable}{sec:oursol} present first results consistent with earlier work and later long-lived solutions in agreement with the observed system. In \secref{sec:discussion} we introduce a concept of evolutionary phases in the lives of wide-orbit planetary systems. These phases could be used to better understand the past and future of a wide-orbit planetary system. We summarise our findings and conclusions in \secref{sec:conclusions}.

%
\section{Simulations}\label{sec:simulations}

\begin{table*}
\caption{Initial conditions of the simulations.}  
\label{tab:simulations}
\centering
\begin{tabular}{l R{9mm} R{9mm} R{9mm} R{10mm} R{10mm} R{10mm} R{10mm} L{55mm}}
\toprule\midrule
Name & \multicolumn{3}{c}{Separation between orbits [$R_H$]} & \multicolumn{4}{c}{Initial semi-major axis [AU]} & Comment \\
& $\Delta _{\mathrm{bc}}$ & $\Delta _{\mathrm{cd}}$ & $\Delta _{\mathrm{de}}$ & $a_{\mathrm{b}}$ & $a_{\mathrm{c}}$ & $a_{\mathrm{d}}$ & $a_{\mathrm{e}}$ & \\
\midrule
\textbf{1} & 4.14 & 3.02 & 3.55 & 67.9 & 38.0 & 24.5 & 14.5 & Face-on assumption. (\secref{sec:sim1})\\ \\
\textbf{2} & 3.47 & 3.31 & 3.90 & 67.9 & 41.9 & 25.8 & 14.5 & Observed inclination, $i = 26\degree$, position angle$ = 64\degree$. (\secref{sec:sim2})\\ \\

\textbf{2$\alpha$} &&&&&&&& Same initial conditions as system~8 in simulation~2. Differ in output rates and thus timestep. (\secref{sec:sim2and4alpha}) \\ \\

\textbf{3} & 3.56 & 3.56 & 3.56 & 68.8 & 41.9 & 24.8 & 14.7 & Optimal inclination, $i = 25\degree$, position angle$ = 42\degree$. (\secref{sec:sim3}) \\ \\
\textbf{4} &&&&&&&& Equal initial orbit separation in $\Delta$. Planet~e fixed at 14.3~AU. (\secref{sec:sim4})\\
$\quad$a & 3.6 & 3.6 & 3.6 & 68.1 & 41.2 & 24.3 & 14.3 \\
$\quad$b & 3.65 & 3.65 & 3.65 & 69.7 & 41.9 & 24.5 & 14.3\\
$\quad$c & 3.7 & 3.7 & 3.7 & 71.2 & 42.5 & 24.7 & 14.3\\
$\quad$d & 3.75 & 3.75 & 3.75 & 72.9 & 43.2 & 24.9 & 14.3\\
$\quad$e & 3.8 & 3.8 & 3.8 & 74.6 & 43.9 & 25.0 & 14.3\\
$\quad$f & 3.85 & 3.85 & 3.85 & 76.3 & 44.6 & 25.2 & 14.3\\
$\quad$g & 3.9 & 3.9 & 3.9 & 78.1 & 45.3 & 25.4 & 14.3\\
$\quad$h & 3.95 & 3.95 & 3.95 & 79.9 & 46.0 & 25.6 & 14.3\\
$\quad$i & 4 & 4 & 4 & 81.8 & 46.7 & 25.8 & 14.3\\ 
$\quad$f\_long & 3.85 & 3.85 & 3.85 & 76.3 & 44.6 & 25.2 & 14.3 & Initialised in the same way as 4f, but not stopping after close encounter. Run time 100~Myr. (\secref{sec:discussion})\\ \\

\textbf{4$\alpha$} &&&&&&&& Same initial conditions as system~94 in simulation~4e. Differ in output rates and thus timestep. (\secref{sec:sim2and4alpha}) \\ \\

\textbf{5} &&&&&&&& Same initial conditions as simulation~4, but planet masses are 4~(b) and 6~(cde)~$M_{\mathrm{Jup}}$. (\secref{sec:sim5and6})\\
$\quad$a & 3.83 & 3.79 & 3.79 & 68.1 & 41.2 & 24.3 & 14.3 \\
$\quad$b & 3.88 & 3.84 & 3.84 & 69.7 & 41.9 & 24.5 & 14.3\\
$\quad$c & 3.93 & 3.90 & 3.90 & 71.2 & 42.5 & 24.7 & 14.3\\
$\quad$d & 3.99 & 3.95 & 3.95 & 72.9 & 43.2 & 24.9 & 14.3\\
$\quad$e & 4.04 & 4.00 & 4.00 & 74.6 & 43.9 & 25.0 & 14.3\\
$\quad$f & 4.09 & 4.05 & 4.05 & 76.3 & 44.6 & 25.2 & 14.3\\
$\quad$g & 4.14 & 4.11 & 4.11 & 78.1 & 45.3 & 25.4 & 14.3\\
$\quad$h & 4.20 & 4.16 & 4.16 & 79.9 & 46.0 & 25.6 & 14.3\\
$\quad$i & 4.25 & 4.21 & 4.21 & 81.8 & 46.7 & 25.8 & 14.3\\ \\

\textbf{6} &&&&&&&& Same initial conditions as simulation~4, but planet masses are 3~(b) and 5~(cde)~$M_{\mathrm{Jup}}$. (\secref{sec:sim5and6})\\
$\quad$a & 4.12 & 4.03 & 4.03 & 68.1 & 41.2 & 24.3 & 14.3 \\
$\quad$b & 4.18 & 4.08 & 4.08 & 69.7 & 41.9 & 24.5 & 14.3\\
$\quad$c & 4.24 & 4.14 & 4.14 & 71.2 & 42.5 & 24.7 & 14.3\\
$\quad$d & 4.29 & 4.20 & 4.20 & 72.9 & 43.2 & 24.9 & 14.3\\
$\quad$e & 4.35 & 4.25 & 4.25 & 74.6 & 43.9 & 25.0 & 14.3\\
$\quad$f & 4.41 & 4.31 & 4.31 & 76.3 & 44.6 & 25.2 & 14.3\\
$\quad$g & 4.46 & 4.36 & 4.36 & 78.1 & 45.3 & 25.4 & 14.3\\
$\quad$h & 4.52 & 4.42 & 4.42 & 79.9 & 46.0 & 25.6 & 14.3\\
$\quad$i & 4.58 & 4.47 & 4.47 & 81.8 & 46.7 & 25.8 & 14.3\\ \\
\bottomrule
\end{tabular}
\end{table*}

\subsection{Orbital parameters}\label{sec:orbital_parameters}

All planets in all our simulations are analogues of the HR~8799~bcde planets and they are initiated on circular, heliocentric orbits. 
 The initial orbit inclination we find by drawing a random inclination between 0\degree\ and 5\degree\ and random longitude of ascending node between 0 and 360\degree\ \citep[$\beta = 5\degree$; see][]{2012ApJ...758...39J}. The initial planet position on the orbit (true anomaly) is randomised. We assign the mass $1.5\; M_{\odot}$ to the star and the planets are given the estimated values of the masses, $5\; M_{\text{Jup}}$ (b) and $7\; M_{\text{Jup}}$ (cde) if nothing else is stated (the masses are decreased in simulations~5 and 6). 
 The planets orbit the barycentre of the system, which is not centred on the star as a result of the high planet
mass. This means that the planetary orbits have a low initial eccentricity as they are initiated on orbits circular around the star. The initial orbital eccentricities vary between different systems as the inclinations, longitude of ascending nodes, and true anomalies are randomised.

We vary the separation between the initial planet orbits (regarding the planets as test particles) in terms of the mutual Hill radius, which is defined as follows:
\begin{equation}\label{eq:RH}
R_H = \frac{r_{p1} + r_{p2}}{2}\left( \frac{M_{p1} + M_{p2}}{3M_{\star}} \right)^{1/3} ,
\end{equation}
where $r$ signifies star-planet separation in the orbital plane and $M$ mass \citep{1996Icar..119..261C}. The indices $p1$ and $p2$ refer to the two planets and $\star$ refers to the star. We then write the separation between the initial, circular orbits  in mutual Hill radii as
\begin{equation}\label{eq:Delta}
\Delta = (r_{p2} - r_{p1})/R_H ,
\end{equation}
where in this case $p2$ refers to the outer planet and $p1$ to the inner planet. For a given planetary system there is a critical $\Delta$ at which in 50\% of the cases the system leads to orbit crossing within a given time \citep{1996Icar..119..261C,2008ApJ...686..580C}, although for any $\Delta$ there is considerable scatter in the times in which particular realisations become unstable.

\subsection{MERCURY6 parameters}

We perform our simulations using the MERCURY6 package \citep{1999MNRAS.304..793C}. We use the Bulirsch-Stoer integrator and a Jacobian coordinate system in the output files. Ejection distance is 11~000~AU and fragmentation from collisions is not allowed. We have redefined a close encounter as when two planets are separated by less than one mutual Hill radius, defined as in \eqref{eq:RH}.

\subsection{Simulation parameters}

In all simulations apart from 4f\_long we stop the integration when a close encounter has occurred or 100~Myr is reached. In 4f\_long we continue running until 100~Myr to investigate the future of multi-planet systems in \secref{sec:discussion}. We use output rates of 10, 100 or 1000~years.  Every simulation contains 100 runs with the difference in initial placement on orbits and the inclinations of the orbits. The simulations we carried out are summarised in \tabref{tab:simulations} and described in detail in \secreftwo{sec:HR8799unstable}{sec:oursol}. We provide the initial conditions of our example system in \tabref{tab:ICs}, but we note that the system is so chaotic that a small change in time-step completely alters the lifetime of the system (see \secref{sec:sim2and4alpha}). A more successful method is to create statistically similar, long-lived systems as explained in this section and described in \secref{sec:oursol} (see also \tabref{sec:simulations}).

Throughout the paper we compare our simulations to the observations of \cite{2011ApJ...729..128C} from 8 October 2009, as these observations have low measurement uncertainties and all four planets are observed simultaneously.

\section{HR~8799 -- unstable as observed}\label{sec:HR8799unstable}

Below we describe in more detail the three simulations 1, 2, and 3 seen in \tabref{tab:simulations}. We use the term stability timescale as the timescale for which the systems in a simulation have not had any close encounters. This timescale we find is roughly the same as the time it takes for the systems to lose a planet. 

\begin{figure}
\centering
\includegraphics[width=\hsize]{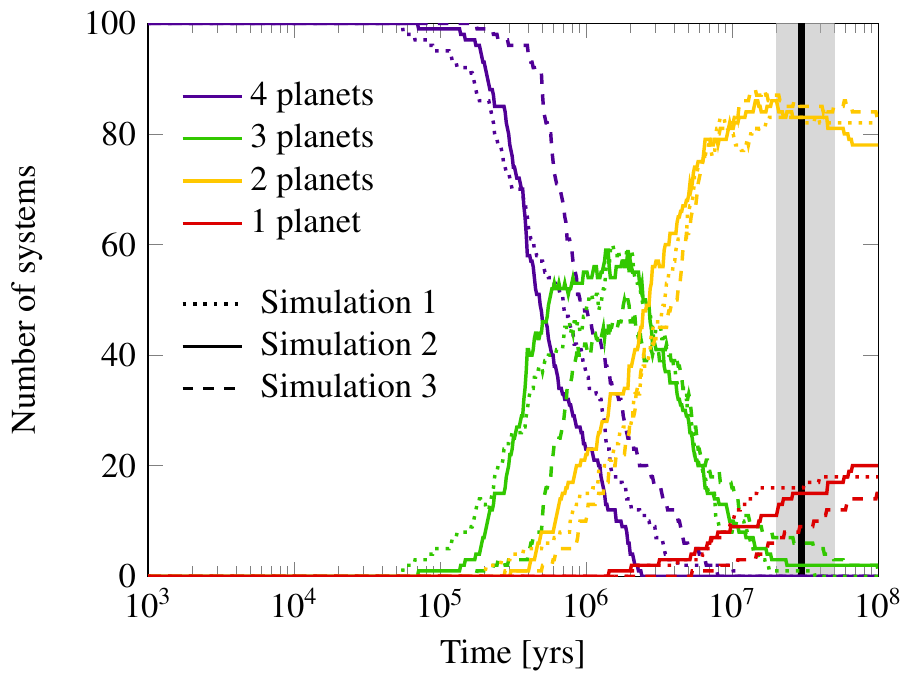}
\caption{Number of planets left in the systems as a function of time. Dotted line corresponds to the face-on assumption (simulation 1), solid line to the inclination inferred from observations (simulation 2), and dashed line to the largest separation achieved from inclination (simulation 3). See \tabref{tab:simulations} and \secref{sec:HR8799unstable} for descriptions about the simulations. The age estimate of HR~8799 ($30^{+20}_{-10}$~Myr) is indicated with a solid line with the grey region corresponding to the uncertainty.}
\label{fig:nbrp_iroptsep_comb}
\end{figure}

In \figref{fig:nbrp_iroptsep_comb} the number of systems with 4, 3, 2, and 1 planets in simulations 1, 2, and 3 is shown as a function of time. The black line indicates the estimated age of HR~8799. The figure shows that none of the systems in simulations 1, 2, or 3 still have four planets when the age of HR~8799 is reached. The figure also shows that $\sim$80\% of the systems have two planets left and $\sim$20\% have one planet left after 100~Myr.

\subsection{Simulation 1: Face-on assumption}\label{sec:sim1}

For simplicity, we begin by assuming the HR~8799 system is seen entirely face-on, i.e. inclination $i = 0\degree$ compared to the plane of the sky. We use the observed on-sky star-planet separations of 67.9 (b), 38.0 (c), 24.5 (d), and 14.5 (e) AU as initial star-planet separations. The initial separations between the orbits are then $\Delta = 4.14$ (b-c), 3.02 (c-d) and 3.55 (d-e)~$R_H$.  The dynamics of the face-on assumption has been tested earlier by \cite{2010ApJ...710.1408F} in the three-planet case and \cite{2010Natur.468.1080M} with all four planets. We confirm their results by finding a median time at which the planetary systems have the first close encounter of $0.10$~Myr and there is no planetary system without close encounters for longer than $9.4$~Myr. We refer to this simulation as simulation 1, see \tabref{tab:simulations} and dotted lines in \figref{fig:nbrp_iroptsep_comb}. Although none of our systems live for the age of HR~8799, we point out here the large range of lifetimes with the longest-lived system surviving 100 times longer than the median.

\subsection{Simulation 2: Inclined system according to observations}\label{sec:sim2}

\begin{figure}
\centering
\includegraphics[scale=1]{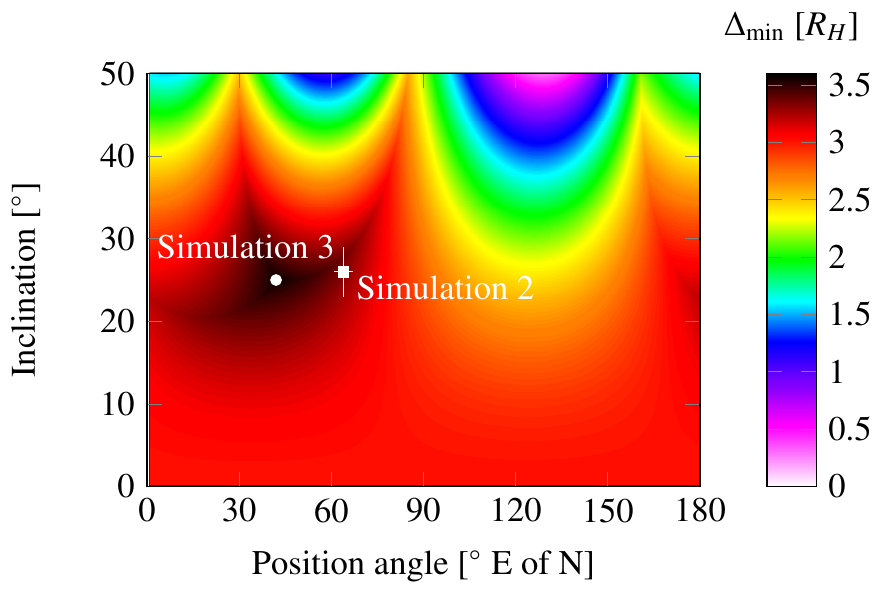}
\caption{Smallest orbit separation, $\Delta _{\min}$, between any planet pair changes with inclination and position angle. We indicate the inclination and position angle of the HR~8799 system measured by \cite{2014ApJ...780...97M} with a white square (simulation~2), which is $i = 26 \pm 3\degree$ and position angle $64 \pm 3\degree$. The largest $\Delta _{\min}$ occurs at $i = 25\degree$ and position angle $42\degree$, when $\Delta _{\min} = 3.56\; R_H$. This value is indicated with a filled, white circle (simulation~3). This plot is based on the observations from \cite{2008Sci...322.1348M,2010Natur.468.1080M}.}
\label{fig:Deltamin_cont}
\end{figure}

The disk of the HR~8799 system is observed to be inclined by $i = 26\degree$ compared to the plane of the sky and  at a position angle of 64\degree E of N \citep[$64 \pm 3\degree$, best fit; G.~Kennedy, private communication; ][]{2014ApJ...780...97M}. Here we assume that the planets and disk are mutually coplanar. These inclination parameters we use also later in simulations 4, 5, and 6. The observed, on-sky separations between planets and star are then 67.9~(b), 41.9~(c), 25.8~(d), and 14.5~(e) AU, which means that planets c and d are located slightly further away from the star compared to the face-on case (simulation 1). The initial separations between the orbits are then $\Delta = 3.47$~(b-c), 3.31~(c-d), and 3.90~(d-e) $R_H$. We refer to this simulation as simulation~2 (see \tabref{tab:simulations}, solid lines in \figref{fig:nbrp_iroptsep_comb} and white square marker in \figref{fig:Deltamin_cont}).

The median time during which the systems have not had any close encounter is in this case $0.094$~Myr, while there are no systems that still have four planets after $9.39$~Myr. Compared to simulation~1 the stability timescale is similar and it remains too short to explain the current state of HR~8799.

\subsection{Simulation 3: Optimal separation from inclination}\label{sec:sim3}

Assume we did not have any system inclination estimate. Which inclination and position angle would then give the largest planetary orbit spacing and thus the longest stability timescale? In \figref{fig:Deltamin_cont} the minimum separation between any two planetary orbits in the system is given for sets of inclination and position angle assuming circular orbits centred on the observed mean \citep[using the observations of][]{2008Sci...322.1348M, 2010Natur.468.1080M}. The largest $\Delta _{\min}$ is found with an inclination of $i = 25\degree$ and a position angle of 42\degree . The separation between the orbits are then $\Delta = 3.56\; R_H$ for all planet pairs, which translates into star-planet separations of 68.8 (b), 41.9 (c), 24.8 (d), and 14.7 (e) AU. We use these optimal separation parameters in simulation 3 (see \tabref{tab:simulations}, dashed lines in \figref{fig:nbrp_iroptsep_comb} and white circle in \figref{fig:Deltamin_cont}). The median time the systems have not had any close encounter is $0.57$~Myr and all systems have had close encounters after $6.85$~Myr.
\figref{fig:nbrp_iroptsep_comb} shows that an increase of $0.25\; R_H$ in $\Delta _{\min}$ results in a factor of $\sim$2 longer stability timescale (compare medians of simulation 2 with simulation 3).

Preferred estimates of the HR~8799 system inclination from a dynamical point of view lie around $20-30\degree$ \citep{2009A&A...503..247R, 2012ApJ...755...38S}, while astroseismological measurements favour inclinations higher than 40\degree assuming that the planetary orbits are aligned with the stellar spin \citep{2011ApJ...728L..20W}.

None of simulations 1, 2 or 3 manage to explain the existence of the HR~8799 system as they simply fall apart before the age of the system is reached.


\section{A stable solution without resonant lock}\label{sec:oursol}

We find long-lived, simulated systems that look like HR~8799 during some time in their evolution. These systems are created with initially wider separation between the orbits than calculated from observations assuming circular orbits. This means that we assume the planets are not moving on completely circular orbits.
 
Our simulated planetary orbits have a low eccentricity of $e \lesssim 0.05$ from the beginning of the integration (for description on how this is set up see \secref{sec:orbital_parameters}). The eccentricity enables planets on more widely separated orbits to look like HR~8799 during parts of the orbits. A wider separation between the orbits in a system can make the system survive longer than the estimated age of HR~8799 as the stability timescale is strongly dependent on orbital separation \citep{2014prpl.conf..787D}.

In this section we present our simulations~4, 5, and 6 (see \tabref{tab:simulations}), which all contain long-lived systems that look like HR~8799. In \secref{sec:sim4} we describe simulation~4 and show an example of a long-lived system that seems to fit the observations of HR~8799. In \secref{sec:stat} we go into detail on how we determine if a simulated system fits the observations. \secref{sec:sim2and4alpha} presents a way of finding even more well-fitting systems. In \secref{sec:MMR} we analyse the simulations in terms of resonant behaviour. \secref{sec:sim5and6} describes simulations~5 and 6, which consider lower planet masses that are still within the estimated range. 

\subsection{Simulation 4: Wider initial orbital separation}\label{sec:sim4}

We test a wider-orbit assumption by simulating systems with the following initial equal separations between the planet orbits: $\Delta = 3.6$, 3.65, 3.7, 3.75, 3.8, 3.85, 3.9, 3.95, and 4~$R_H$. We place planet~e at an initial star-planet separation of 14.3~AU, which then fixes the initial semi-major axes of the other planets. The planet masses are set to the nominal masses $M = $ 5~$M_{\rm Jup}$ (b) and 7~$M_{\rm Jup}$ (cde) \citep{2010Natur.468.1080M}. These simulations are called simulation 4a--i; see \tabref{tab:simulations}. Simulation~4 thus contains 900 runs compared to simulations~1, 2, and 3, which only contain 100 runs each.
It makes sense to talk about simulations 4a--i for the initialisation of the runs, but not in terms of the architecture of the outcome. The measured minimum orbit separation $\Delta _{\min}$ for each system ranges between 3.6 and 4~$R_H$ in almost each of simulations 4a--i, depending on the orbital eccentricities. Therefore we bin the simulations~4 with $\Delta _{\min}$, regardless of the initial set-up. We make nine bins with each 100~runs. These bins contain different ranges of orbit separations ($\Delta _{\min}$) and thus represent slightly different architectures of the planetary system.

\begin{figure}
\centering
\includegraphics[width=\hsize]{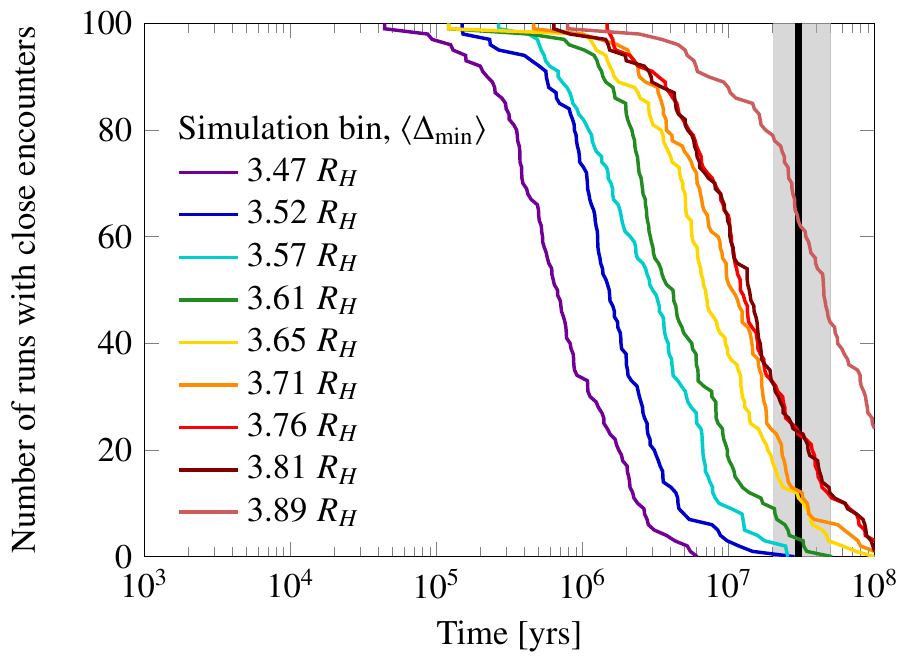}
\caption{Number of systems without close encounters shown with time. The different coloured lines correspond to bins of each 100~systems in simulation~4, sorted in measured minimum orbit separations ($\Delta _{\min}$). The legend indicates the bins and their systems' mean value of $\Delta _{\min}$. The black line and the grey zone show the age estimate with uncertainty of HR~8799.}
\label{fig:4ness_Delta}
\end{figure}

\figref{fig:4ness_Delta} shows the time during which the systems in the simulation~4 bins have not had any close encounter. The figure shows the trend of stability timescale with minimum orbit separation, $\Delta _{\min}$. Compared to the systems in \figref{fig:nbrp_iroptsep_comb} a substantial fraction of systems in \figref{fig:4ness_Delta} have not experienced any close encounter at the estimated age of HR~8799 ($\sim$30~Myr). \cite{2008ApJ...686..580C} find that a larger number of mutual Hill radii are needed to keep their planetary systems stable compared to what we find. For various reasons (e.g. our higher planet masses, lower initial eccentricities, and different stability criterion), we consider both results valid and our results not contradicting the results of \cite{2008ApJ...686..580C}. We see that at any given separation there is a wide range of stability times, a fact to which we return later.

\begin{figure*}
\centering
\begin{subfigure}[t]{.45\textwidth}
\centering
\includegraphics[scale=1]{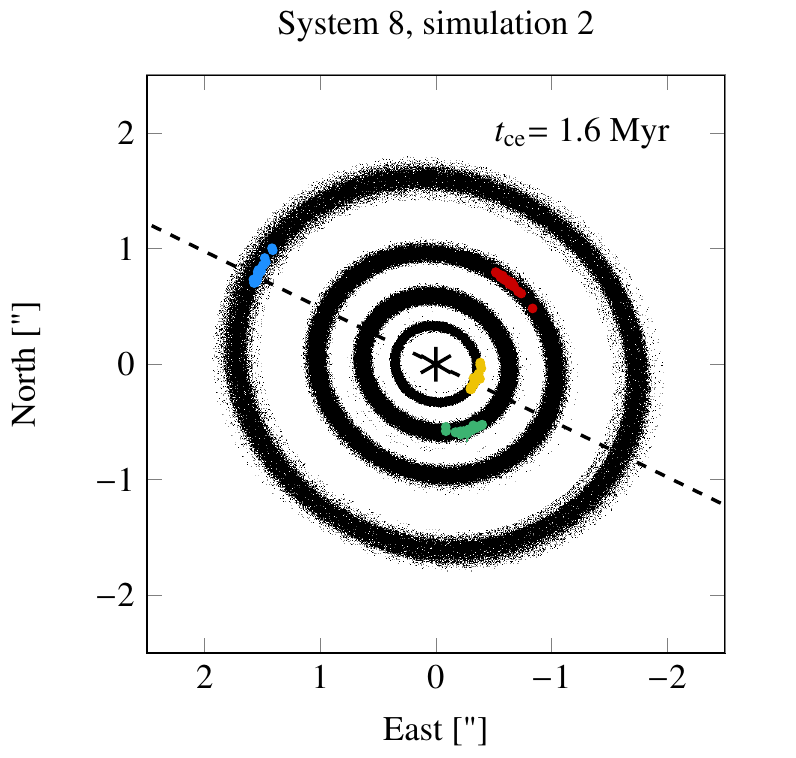}
\caption{System 8 in simulation 2 shown together with the observations of the HR~8799 planetary system. The system seems to fit the observations well, but suffers a close encounter at an age of 1.6~Myr between planet d and e. In the figure we show the positions of the simulated planets every 10 years (black dots).}
\label{fig:sys8_sim2}
\end{subfigure}
\hspace{5mm}
\begin{subfigure}[t]{.45\textwidth}
\centering
\includegraphics[scale=1]{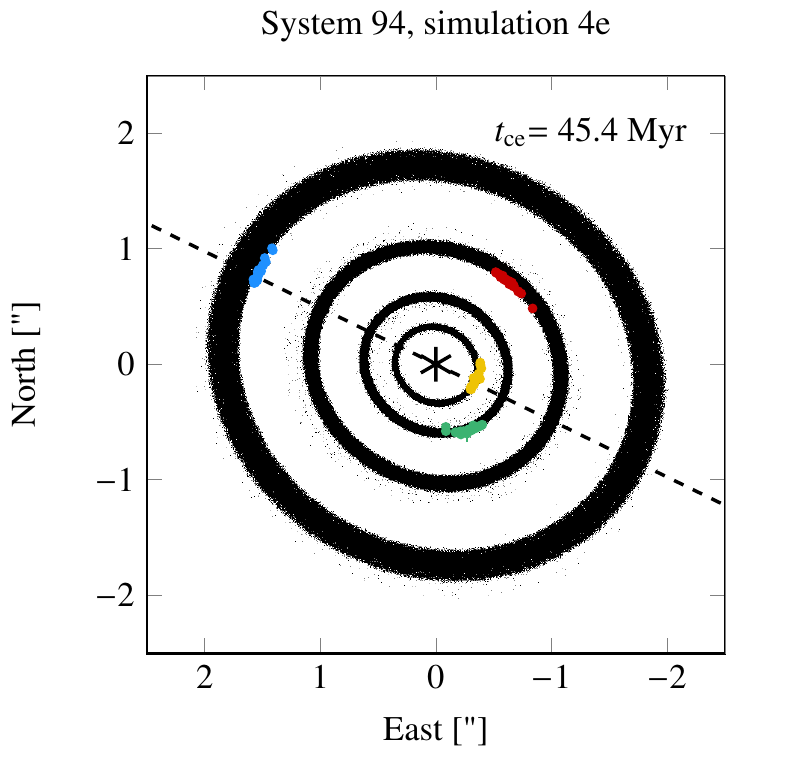}
\caption{System 94 in simulation~4e shown together with the observations of the HR~8799 planetary system. This system seems to fit the observations well until a close encounter occurs between planets d and e at an age of 45.4~Myr. We use this system as an example of a long-lived system without strong resonances that also look like HR~8799. In the figure we show the positions of the simulated planets every 100 years (black dots).}
\label{fig:sys94_sim4e}
\end{subfigure}
\caption{Comparison between a system in simulation~2 and a system in simulation~4e. Both systems seem to fit the observed positions of the planets, but only the system in simulation~4e is stable for longer than the estimated age of HR~8799. This is an on-sky projection of the simulations assuming a distance of 39.4~pc \citep{2007A&A...474..653V} and inclination 26\degree\ \citep[position angle 64\degree ,][]{2014ApJ...780...97M}. The axis around which the system is inclined is indicated with a dashed line. The position of the star is indicated with an asterisk. }
\label{fig:tyre}
\end{figure*}

Do the surviving systems in simulation~4 look like HR~8799? In \figref{fig:tyre} we compare a system from simulation~2 (left panel, system 8) with a system from simulation~4e (right panel, system 94). Both runs seem to fit the observations, but the system from simulation~4 survives over 45~Myr because of the slightly more widely separated planetary orbits. We want to stress that system~94, simulation~4e is not a special system, but seems normal compared to the other systems in simulation~4. As we explain in \secref{sec:stat} we find 12 systems fitting the observations in simulation~4. If we consider a distance to the HR~8799 system $1\sigma$ away from the mean of the estimate ($39.4+1.1$~pc $= 40.5$~pc) we almost double the number to 22 fitting systems in simulation~4. How we determine if a system fits the observations is discussed in detail in \secref{sec:stat}.

\begin{figure}
\centering
\includegraphics[scale=1]{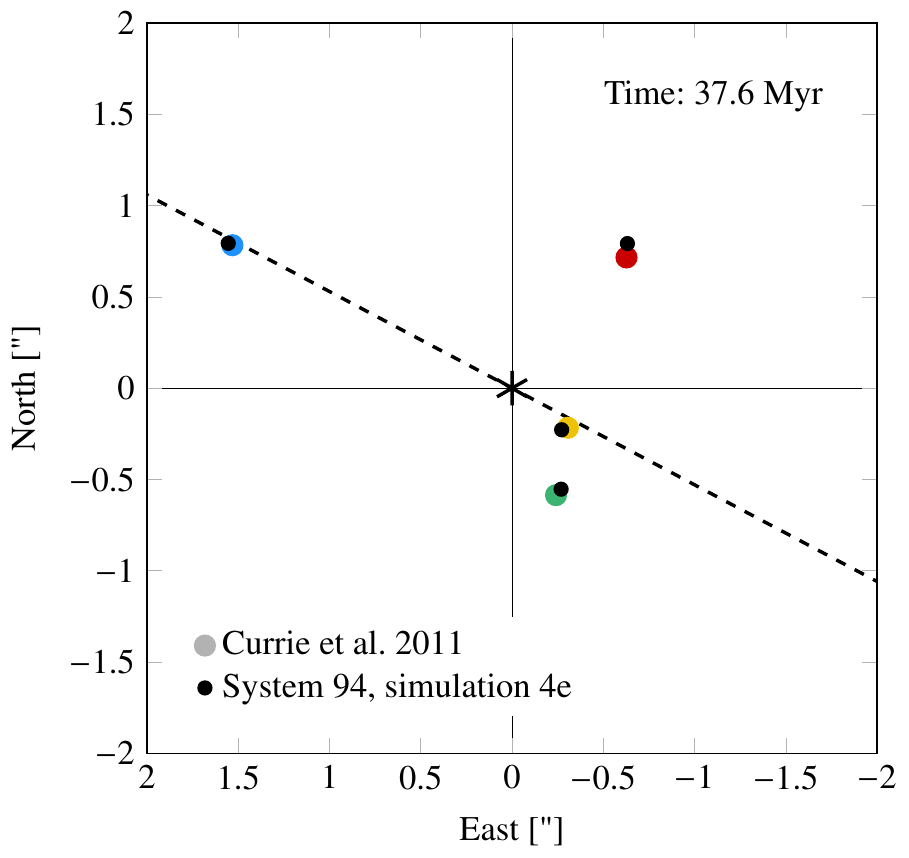}
\caption{System 94, simulation 4e at 37.6~Myr (small black dots) compared to the observations of \cite{2011ApJ...729..128C} (coloured, larger dots). The axis around which the system is inclined is shown with a dashed line \citep[position angle 64\degree\ and inclination 26\degree ,][]{2014ApJ...780...97M}. The position of the star is indicated with an asterisk.}
\label{fig:match}
\end{figure}

To demonstrate the similarity of our example system, system 94, to the observed HR~8799 system, we show a snapshot of the system at 37.6~Myr in \figref{fig:match}.

\subsection{Determining whether a system fits the observations}\label{sec:stat}

In this section we describe in detail how we determine how each simulated system fits the observations. We have chosen to compare the simulated systems with the observations taken on 8 October 2009 by \cite{2011ApJ...729..128C} because the planets are observed simultaneously during this observation and the measurement uncertainties are comparably small. 

\begin{figure*}
\centering
\includegraphics[width=\textwidth]{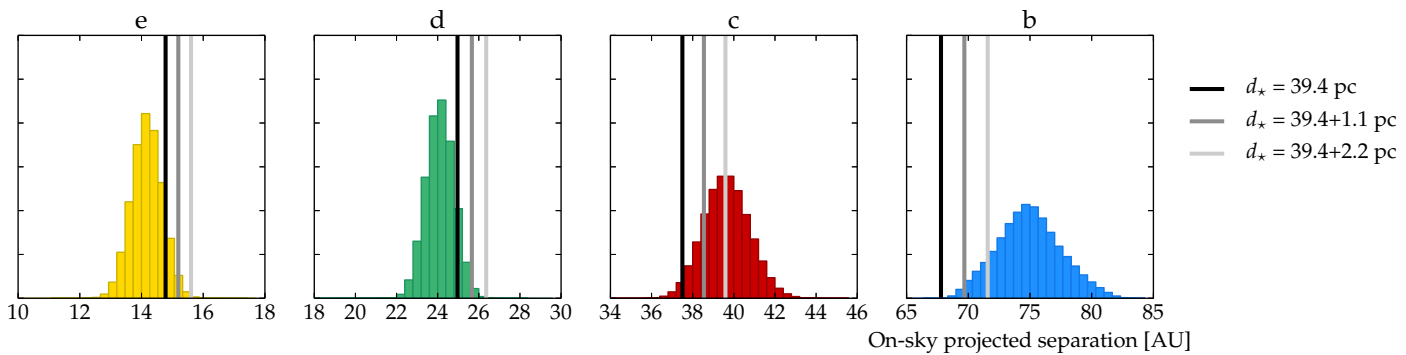}
\caption{Comparison of simulations to observed planet positions for planet e, d, c, and b subsequently. The histograms show the star-planet distance in our system 94, simulation 4e, after having moved all output to the observed position angle of the planet, bootstrapped a data set, and added a normally distributed error to the simulation data. We show, in black, dark grey, and light grey lines, the observed on-sky star-planet distances from \cite{2011ApJ...729..128C} assuming the mean Earth-HR~8799 distance (39.4~pc) and adding 1 and 2$\sigma$ of this distance estimate \citep{2007A&A...474..653V}. This procedure is described in detail in \secref{sec:stat}.}
\label{fig:hist_stat}
\end{figure*}

We start by checking whether the system has a longer stability timescale than the estimated age of HR~8799 
\citep[30~Myr,][]{2010Natur.468.1080M}. If that is the case, we proceed with checking whether the system looks like HR~8799. We check whether the architecture of the simulated system is similar to the architecture of HR~8799 in the following way:

\begin{enumerate}
\item For every output, move each planet along its orbit to the observed position angle. This is done because it is highly unlikely that we find a single output where all four planets happen to be located within the uncertainties of the observations. The space the observations take on compared to the orbit is about 0.04~AU/200~AU = $2 \times 10^{-4}$. The chance of then catching a point in time when all four planets are within the observations would be $(2 \times 10^{-4})^{3} = 8 \times 10^{-12}$ for a very well-fitting system (fixing for example b to always be at the observed position angle). To catch one output where all planets fit simultaneously would therefore require $\sim 10^{11}$ outputs  -- a number about three orders of magnitude higher than our output rate. As during the simulation the planets are at some point located at exactly the observed position angle, but we do not catch that moment, it is validated to move the planets along their orbits in this way to really measure how close they come to the observations. \\ 
\item Draw one million random star-planet separations from the distributions created in step number 1 for each planet. This is carried out to create a new distribution of star-planet separations with a method called bootstrapping with resampling. \\ 
\item To simulate the uncertainty of the observation, draw a normally distributed error to each star-planet separation in the bootstrapped distribution using the standard deviation of the observations. We show these distributions for the simulated planets in system 94, simulation~4e in \figref{fig:hist_stat}.\\
\item Measure for each planet the quantile for the observed star-planet separation in the created distribution. The quantile is the fraction of outputs residing to the left of the observed star-planet separation. The star-planet separations are indicated with black, dark grey, and light grey lines in \figref{fig:hist_stat} for Earth-HR~8799 distances of 39.4~pc, then $+1$ and $+2\sigma$ on the distance estimate.
\end{enumerate}
The quantiles are the measures for how well the observed planet position and the simulated planet positions fit. Quantiles close to 0.5 would indicate a very good fit. We assess systems with quantiles between $10^{-4}$ and $1 - 10^{-4}$ as systems fitting to the observations. A low limit for the quantiles means that we allow systems that rarely are found with the architecture of the HR~8799 system to also count as fitting systems. Our example system, system 94, simulation~4e actually has a lower number for the quantile for planet~b and does not count as a fitting system.

A visualisation of the assessment procedure can be seen in \figref{fig:hist_stat} where we show, for system 94, simulation~4e, the star-planet distributions plotted together with the star-planet distances inferred from the Earth-HR~8799 distances 39.4, 40.5, and 41.6~pc (corresponding to the mean of the estimate, then one and two standard deviations away from the mean). The quantiles for system 94, simulation~4e are $q_{\rm e} = 0.89$, $q_{\rm d} = 0.89$, $q_{\rm c} = 0.02$ and $q_{\rm b} = 8.7 \times 10^{-5}$ using the mean of the Earth-HR~8799 distance estimate. If we consider the distance one standard deviation away from the estimate (40.5~pc to HR~8799), however, we find the quantiles $q_{\rm e} = 0.98$, $q_{\rm d} = 0.99$, $q_{\rm c} = 0.16$ and $q_{\rm b} = 0.01$. At this distance we consider the system 94, simulation~4e to fit the observations well. To not miss systems such as system 94, simulation~4e in our analysis, we consider two values of the Earth-HR~8799 mean of the estimate distance (39.4~pc) and the distance one standard deviation away from the mean of the estimate (40.5~pc).

\begin{figure}
\centering
\includegraphics[scale=1]{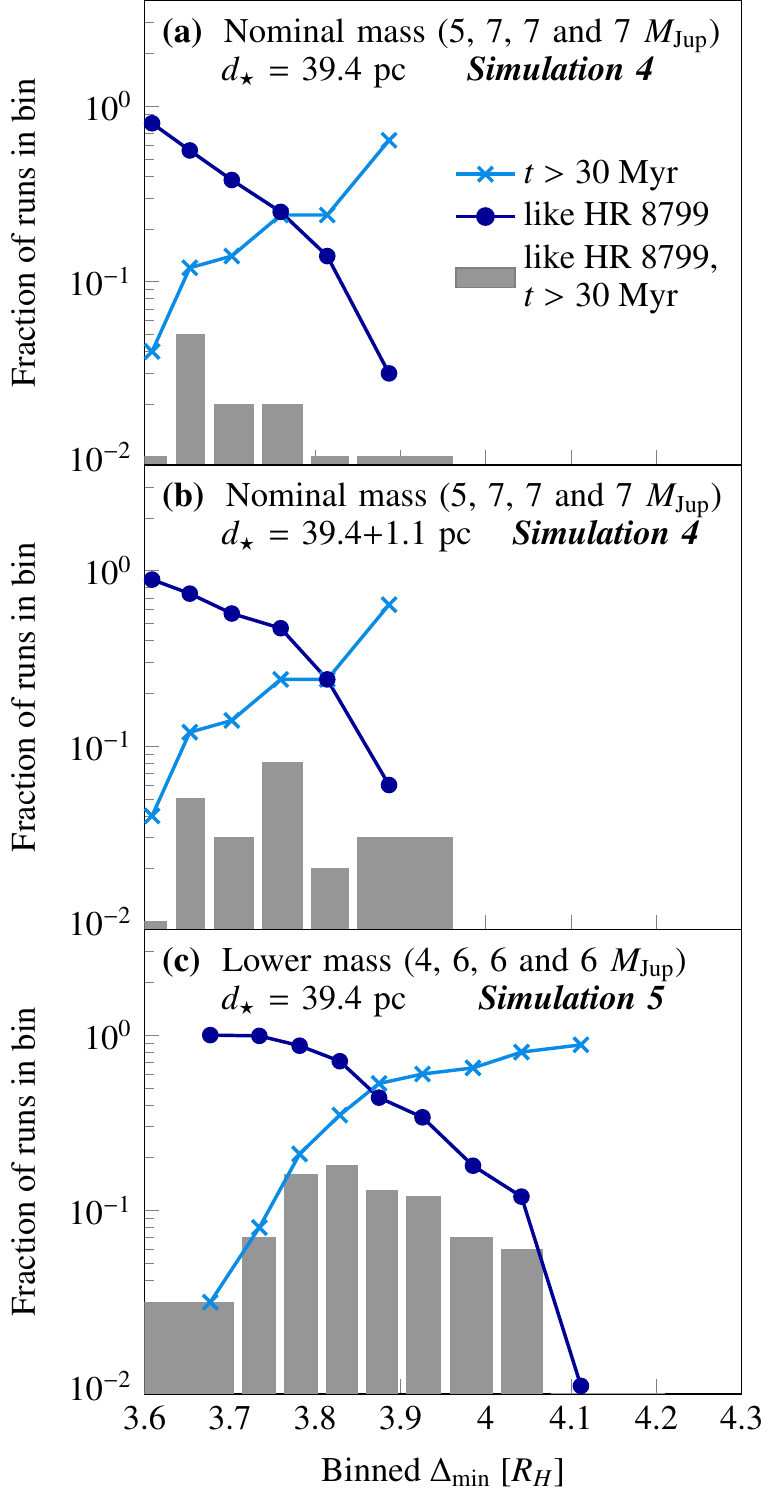}
\caption{We show how many of our systems in simulation~4 (top and middle panels) and simulation~5 (bottom panel) fit the observations. We binned the simulations in ranges of minimum orbit separations, $\Delta _{\min}$, 100 systems in each bin. The light blue shows the fraction of the systems in the bin that have a stability timescale longer than the estimated age of HR~8799. The dark blue curve shows the fraction of the systems in the bin that have architectures fitting to that of HR~8799. We determine whether a system fits through the procedure described in \secref{sec:stat}. The grey bars indicate the fraction of systems in each bin that fulfil both criteria. The top panel shows the results from simulation~4 using the mean of the Earth-HR~8799 distance estimate (39.4~pc), the middle panel shows the results from simulation~4 using a one standard deviation longer Earth-HR~8799 distance (40.5~pc), and the bottom panel shows the results from simulation~5, which has slightly lower planet masses, but is still within the estimated range.}
\label{fig:frac8799surv}
\end{figure}

We summarise our findings of simulation~4 in the top two panels of \figref{fig:frac8799surv}. Again we have binned the simulation~4 with measured minimum orbit separation, $\Delta _{\min}$, 100 systems in each bin. This makes in total 9 bins and 900~systems. We plot in light blue the fraction of the systems in each bin that have a stability timescale longer than the estimated age of HR~8799. In dark blue we show the fraction of systems in each bin that fit the observations of the HR~8799 system according to the above described method. The grey bars show the fraction of systems in each bin that fulfil both criteria and thus fit the HR~8799 system. We find for the nominal Earth-HR~8799 distance (39.4~pc), in the best case (where $\Delta _{\min} = 3.64-3.67$) 5 systems of 100 that fit the observations. In the entire range of orbit separations, we find 12 of 900 systems that fit the observations (see \figref{fig:frac8799surv}a). For the Earth-HR~8799 distance of 40.5~pc we find, in the best case (where $\Delta _{\min} = 3.74 - 3.79$) 8 systems of 100 fitting the observations and 22 of 900 when including all orbit separation ranges (see \figref{fig:frac8799surv}b). 

In simulation~4 we create many varieties of architectures in the planetary systems that they would fit many ranges of inclinations and position angles. Therefore we are not concerned by the exact values of the inclination and position angle that we apply \citep[from][]{2014ApJ...780...97M}. 

\subsection{Simulations 2$\alpha$ and 4$\alpha$: Verification}\label{sec:sim2and4alpha}

\begin{figure}
\centering
\includegraphics[scale=1]{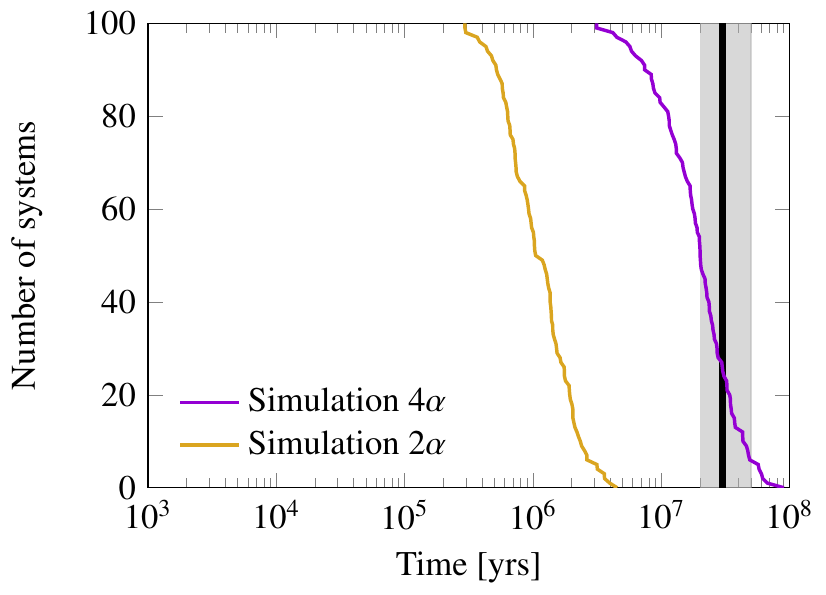}
\caption{Comparison between simulations 4$\alpha$ (violet) and 2$\alpha$ (yellow). The lines show the number of systems in the simulation that have not experienced a close encounter, plotted with time. The slightly wider orbits in simulation~4$\alpha$ compared to simulation~2$\alpha$ makes the systems close encounter time shift about an order of magnitude. The black line and the grey zone show the age estimate and uncertainty of HR~8799.}
\label{fig:copsims}
\end{figure}

SSystem~94, simulation~4e (see e.g. \figref{fig:sys94_sim4e}) is not a special system, but there are many evolutions leading to systems that fit the observations of HR~8799. To show this, we simulate 100 copies of system~94, simulation~4e with the only difference being the simulation output rate, which changes the timestep slightly. This very small timestep variation changes the fate of the systems in$\sim 10^6$~years because they are chaotic. The systems have similar architecture, but with different evolution. We call them simulation~4$\alpha$.

Because our simulated systems are very chaotic it is likely that an integration made with our exact initial conditions would evolve differently from that which we present here. A small change in timestep alters the evolution on a timecale of $10^6$~yrs and our interesting systems have a stability timescale much longer than that. 
Therefore, we note that our initial conditions in \tabref{tab:ICs} are likely to not reproduce our system 94, simulation 4e. Instead, we present our recipes on how to create similar systems.

We also create simulation~2$\alpha$ from system~8, simulation~2, using the same procedure as when creating simulation~4$\alpha$. \figref{fig:copsims} shows with time the number of systems in simulations~$2\alpha$ and $4\alpha$ that have not experienced any close encounter. The slopes of the curves show that the systems indeed have different evolutions despite their identical initial conditions. The location of the curves in the diagram shows that the stability timescale of simulation~$4\alpha$ is about one order of magnitude longer than the stability timescale of simulation~$2\alpha$. This leads to 26 systems in simulation~$4\alpha$ surviving longer than the estimated age of HR~8799 and 10 of these also matching the observations of HR~8799.

\subsection{Mean motion resonance analysis}\label{sec:MMR}

In this section we investigate whether resonances affect our systems in simulation~4. Stabilisation because the system is deep in a mean-motion resonance is a common explanation to the survival of the HR~8799 system \citep[see e.g.][]{2010ApJ...710.1408F, 2009MNRAS.397L..16G, 2009A&A...503..247R, 2010ApJ...721L.199M, 2011ApJ...729..128C, 2011ApJ...741...55S, 2013A&A...549A..52E, 2011ApJ...741...55S, 2014MNRAS.440.3140G}. In the following we show that we simulate systems that fit the HR~8799 system without being stabilised by strong resonant lock.

We consider resonant angles correlated to the Laplace resonances, 2:1, 4:2:1, and 8:4:2:1 \citep{1999ssd..book.....M}. We also consider angles related to the secular motion. All these angles are calculated using the mean longitude ($\lambda$) and longitude of periastron ($\varpi$), which are calculated from the longitude of ascending node ($\Omega$), argument of periastron ($\omega$), and mean anomaly ($M$) for each planet in the following way:
\begin{eqnarray}
\lambda &=& \Omega + \omega + M  \\ 
\varpi &=& \Omega + \omega 
.\end{eqnarray}
The resonant angles we consider of the 2:1 resonance are
\begin{eqnarray}
\phi _{1, \; 2:1} =& 2\lambda _{\rm outer} - \lambda _{\rm inner} - \varpi _{\rm outer} \label{eq:phi1_21} \\
\phi _{2, \; 2:1} =& 2\lambda _{\rm outer} - \lambda _{\rm inner} - \varpi _{\rm inner} \label{eq:phi2_21}
,\end{eqnarray}
which for the three planet pairs (b-c, c-d and d-e) result in six angles to consider.

For the 4:2:1 resonance we consider the angle
\begin{equation}\label{eq:phi1_421}
\phi _{1, \; 4:2:1} = \lambda _{\rm inner} - 3\lambda _{\rm middle} + 2\lambda _{\rm outer}
,\end{equation}
which results in two angles as there are two sets of three-planet systems; b-c-d and c-d-e.

For the 8:4:2:1 resonance we take into account the four angles as follows:
\begin{eqnarray}
\phi _{1, \; 8:4:2:1} =& \lambda _{\rm e} - 2\lambda _{\rm d} - \lambda _{\rm c} + 2\lambda _{\rm b}  \\ \label{eq:phi1_8421}
\phi _{2, \; 8:4:2:1} =& 2\lambda _{\rm e} - 5\lambda _{\rm d} + \lambda _{\rm c} + 2\lambda _{\rm b} \\
\phi _{3, \; 8:4:2:1} =& \lambda _{\rm e} - \lambda _{\rm d} - 4\lambda _{\rm c} + 4\lambda _{\rm b} \\
\phi _{4, \; 8:4:2:1} =& \lambda _{\rm e} - 4\lambda _{\rm d} + 5\lambda _{\rm c} - 2\lambda _{\rm b}
,\end{eqnarray}
which we now label with the planet names. 

We also look at the difference in longitude of periastron, which is an angle related to secular motion. It is defined as
\begin{equation}\label{eq:varphi_sec}
\varphi  = \varpi _{\rm inner} - \varpi _{\rm outer}
,\end{equation}
which leads to three angles from the three planet pairs (b-c, c-d and d-e).

\begin{figure*}
\centering
\includegraphics[scale=1]{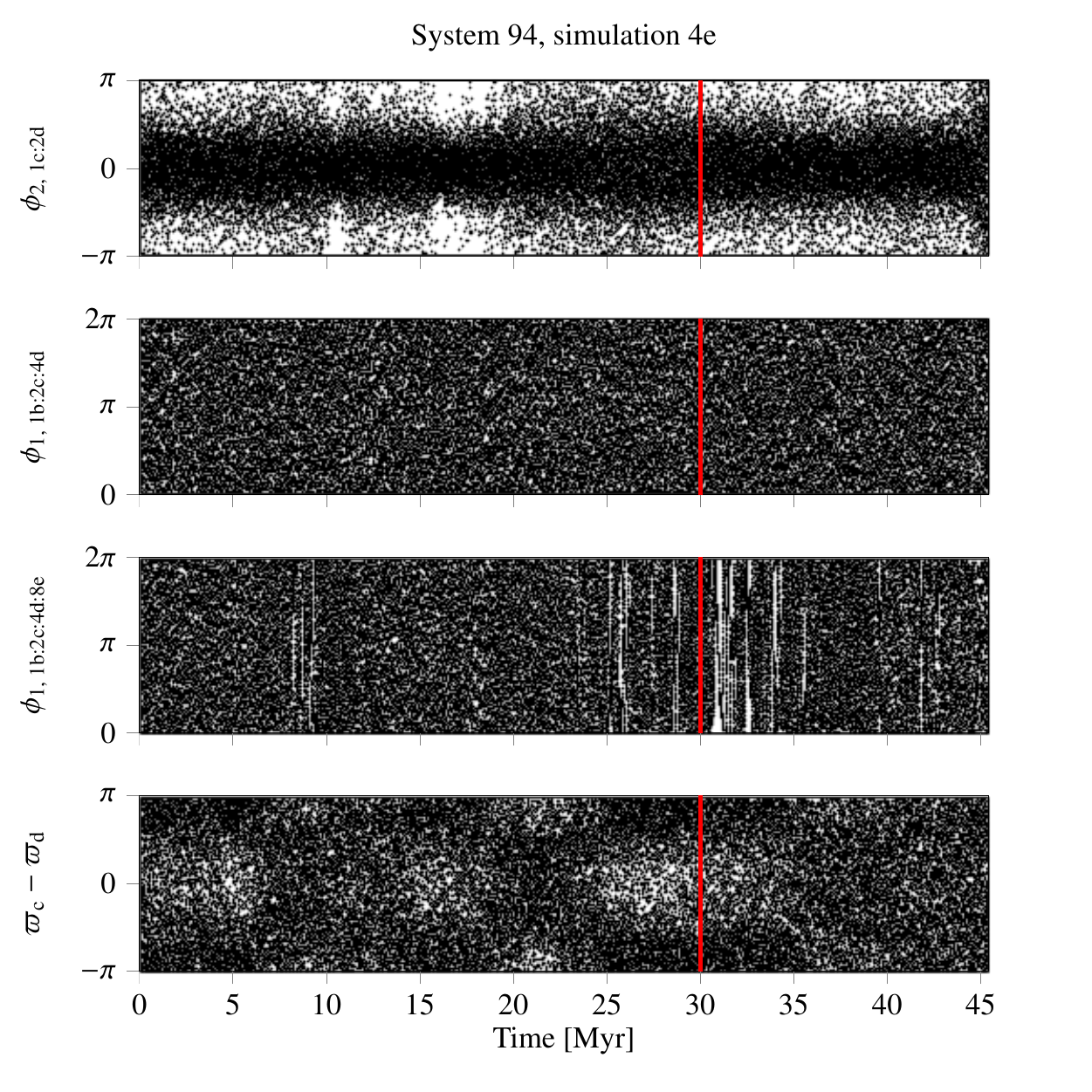}
\caption{Evolution of from top to bottom the resonant angles $\phi _{2, \; {\rm1c:2d}}$, $\phi _{1, \; {\rm 1b:2c:4d}}$, $\phi _{1, \; {\rm 1b:2c:4d:8e}}$ and $\varpi _{\rm c} - \varpi _{\rm d}$ for system 94, simulation~4e. The black dots show the output from our simulation, sampled every 100~years. In the top panel (evolution of the $\phi _{2, \; {\rm1c:2d}}$) the output clusters around 0~rad, which is a sign of weak resonant behaviour. The system alternates between libration and circulation in this resonant angle, which is a different behaviour than what is seen in systems affected by strong resonance in which all output would be narrowly confined about 0~rad. None of the bottom three panels show signs of resonant libration, unlike in the cases considered by \cite{2009MNRAS.397L..16G, 2014MNRAS.440.3140G}. The red line indicates the estimated age of HR~8799.}
\label{fig:res_angles_94_4e}
\end{figure*}

In \figref{fig:res_angles_94_4e} we show the evolution of four of the above angles in system~94, simulation~4e. We want to stress that system 94, simulation~4e is a system similar to many of our other systems in the resonant behaviour seen in \figref{fig:res_angles_94_4e} also. The evolution of the angles in our panels 2, 3, and 4 in \figref{fig:res_angles_94_4e} are also shown in the mean motion resonance analysis of \cite{2009MNRAS.397L..16G} (their Figs.~4, middle panel, and 6, top panel) and \cite{2014MNRAS.440.3140G} (their Fig.~1, bottom right panel). 
The previously presented solutions \citep[see e.g.][]{2009MNRAS.397L..16G, 2010ApJ...710.1408F, 2014MNRAS.440.3140G} all show small-amplitude librations of one or more resonant angles. In contrast, our solution shows at most transitions between large-amplitude libration and circulation. This configuration does not guarantee the indefinite protection from close encounters that small-amplitude librations provide. 

\begin{figure*}
\centering
\includegraphics[width=\hsize]{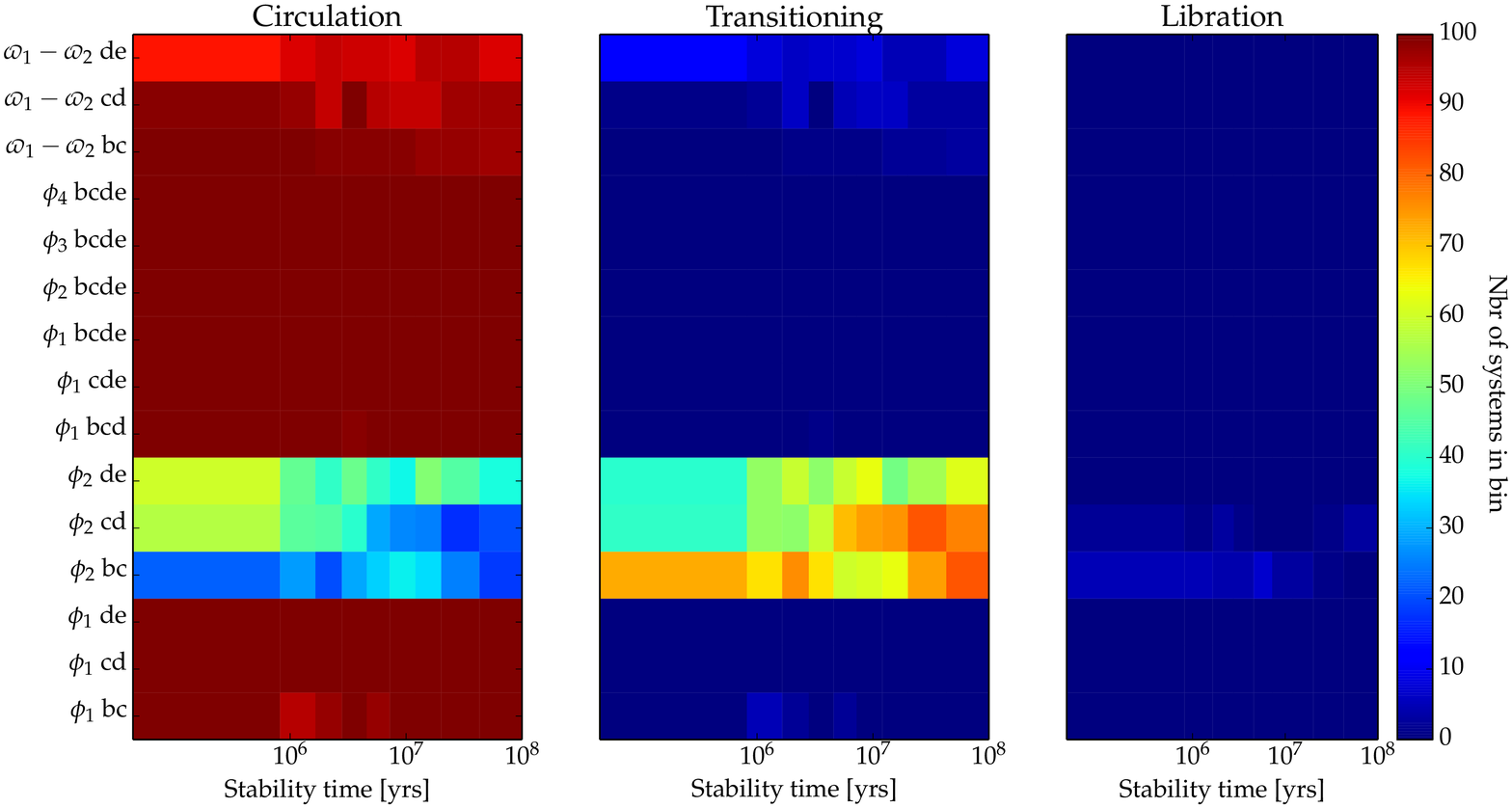}
\caption{Resonant behaviour of our simulation~4, binned with stability time ($x$-axis) and 100 systems per bin. Colour shows for different resonant angles ($y$-axis) how many systems in the bin are circulating (left panel), librating (right panel) or transitioning between a resonant and non-resonant state (middle panel). We consider the angles in Eqs.~\ref{eq:phi1_21}--\ref{eq:varphi_sec} (see \secref{sec:MMR}). Our stability times in simulation~4 stretch between 0.04~Myr up to stable systems after 100~Myr. We see a resonant behaviour in the 2:1 resonant angles where it is common that the systems transition between libration and circulation. However, there seems to be no dependence on stability time with resonances.}
\label{fig:pcolor_MMR}
\end{figure*}

To show the overall behaviour of our simulation~4 we bin the systems by stability time into 9 bins, each containing 100 systems. For each system we assess whether the resonant angles presented in Eqs.~\ref{eq:phi1_21}--\ref{eq:varphi_sec} are circulating, librating, or falling in and out of resonance.  We visualise the resonant behaviour of our simulation~4 in \figref{fig:pcolor_MMR}. \figref{fig:pcolor_MMR} shows no trend of longer stability time with librating angles (right panel) and no clear trend of stability time angles transitioning between librational and circulatory behaviour. We infer from this that our systems are not being stabilised by tight resonant lock. However, the separatrix-crossing behaviour we observe in some systems is characteristic of chaotic evolution driven by multiple resonances \citep{1979PhR....52..263C}. Indeed, this may offer an explanation as to why there is a long-lived tail of surviving systems with no consistent resonant protection: \cite{2010ApJ...712..819S} describe how some trajectories stick for extended periods at the edge of the chaotic region of phase space.

\subsection{Simulations 5 and 6: Lower planet masses}\label{sec:sim5and6}

Another way to increase the stability timescale of a planetary system is to decrease the planet masses. In terms of $\Delta$ (\eqreftwo{eq:RH}{eq:Delta}), lowering planet masses widens the orbit separations and thus increases the stability timescale.

We use the exact same initial conditions for simulation~4 and lower the planet masses to create simulation~5 and 6 (see \tabref{tab:simulations}). Simulation~5 has planet masses 4 (b) and 6 (cde) $M_{\rm Jup}$ and simulation~6 has planet masses 3 (b) and 5 (cde) $M_{\rm Jup}$. We want to point out that these planet masses are still comfortably within the estimated masses from cooling models \citep[$5\pm 2\;M_{\rm Jup}$ (b) and $7^{+3}_{-2}\;M_{\rm Jup}$ (cde);][]{2008Sci...322.1348M, 2010Natur.468.1080M}. 

\begin{figure}
\centering
\includegraphics[scale=1]{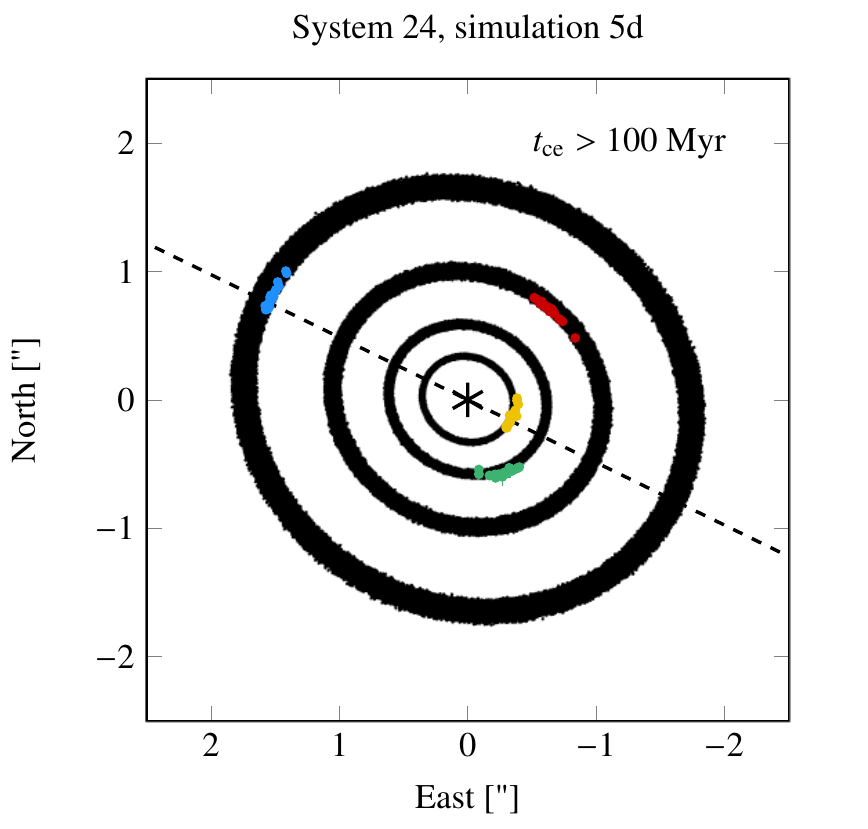}
\caption{System~24 in simulation 5d (using lower planet masses of 4~$M_{\rm Jup}$ for planet e and 6~$M_{\rm Jup}$ for planets b, c, and d). Black dots are our output data points assuming the inclination and position angle of 26\degree and 64\degree from \cite{2014ApJ...780...97M}. The coloured dots are the observations of the HR~8799 planetary system. 
The system matches the data and does not experience any close encounter during our 100~Myr integration time.}
\label{fig:tyre_245d}
\end{figure}

\begin{figure*}
\centering
\includegraphics[width=\textwidth]{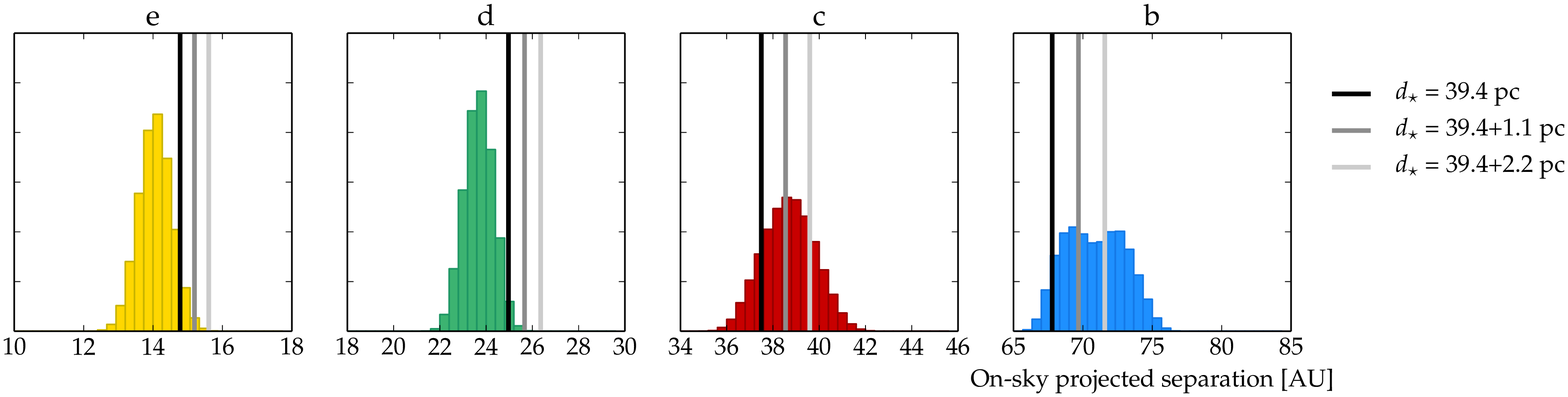}
\caption{Same as \figref{fig:hist_stat} but for the system 24, simulation 5d (planet masses are lower than nominal, $M = $ 4~$M_{\rm Jup}$(e), and 6~$M_{\rm Jup}$ (bcd)).}
\label{fig:hist_245d}
\end{figure*}

An example of a system that fits from simulation~5 is system 24, simulation~5d seen in \figreftwo{fig:tyre_245d}{fig:hist_245d}. The system does not experience any close encounter during the 100~Myr integration time and we measure the quantiles to be $q_{\rm e}= 0.93$, $q_{\rm d} = 0.98$, $q_{\rm c} = 0.15$ and $q_{\rm b} =0.07$ at the nominal Earth-HR~8799 distance 39.4~pc. We consider this system to be a good fit to the observations of HR~8799. The system is not stabilised by low-amplitude librational behaviour.

Using the technique described in \secref{sec:stat} we find that for simulation~5, in the best case (where $\Delta _{\min} = 3.81 - 3.85$) 18 systems of 100 fit the observations. Looking at a broader range of orbit separations in total 82 out of 900 systems fit the observations (see \figref{fig:frac8799surv}c). Considering these 1~$M_{\rm Jup}$ lower planet masses in simulation~5 the architecture of HR~8799 is becoming common in our systems also after 30~Myr. The boost of fitting systems in simulation~5 is clearly seen in \figref{fig:frac8799surv} when comparing the bottom panel (simulation~5) with the top two panels (simulation~4).

We include simulation~6 to prove the point that lower planet masses implies much longer stability timescale. In simulation 6 we find 64 fitting systems of 100 in the best case (when $\Delta _{\min} = 3.98 - 4.02$) and 241 of 900 when considering all orbit separation ranges.

Considering a slightly larger Earth-HR~8799 distance increased the number of fitting systems with a factor of two \secref{sec:sim4}. Considering slightly lower planet masses boosts the number of fitting systems by a factor of ten. 


\section{Evolutionary phases}\label{sec:discussion}

In this section we use simulation~4f\_long (see \tabref{tab:simulations}) to investigate the evolution of wide-orbit planetary systems. Simulation~4f\_long is created in the same way as simulation~4f, but the integrations are not stopped after a close encounter has occurred and therefore always continue up to 100~Myr. 

We find three distinct phases of dynamical evolution in our wide-orbit planetary systems. We call the phases A, B, and C and they correspond to the time before the first close encounter between planets (A), the time between the first and last close encounter (B), and the time after the last close encounter (C). We define a close encounter between planets to be whenever two planets enter their mutual Hill radius, see \eqref{eq:RH}. We can use these evolutionary phases to predict what will happen and what has happened to the HR~8799 system and systems similar to it.

\begin{figure*}
\centering
\includegraphics[scale=1]{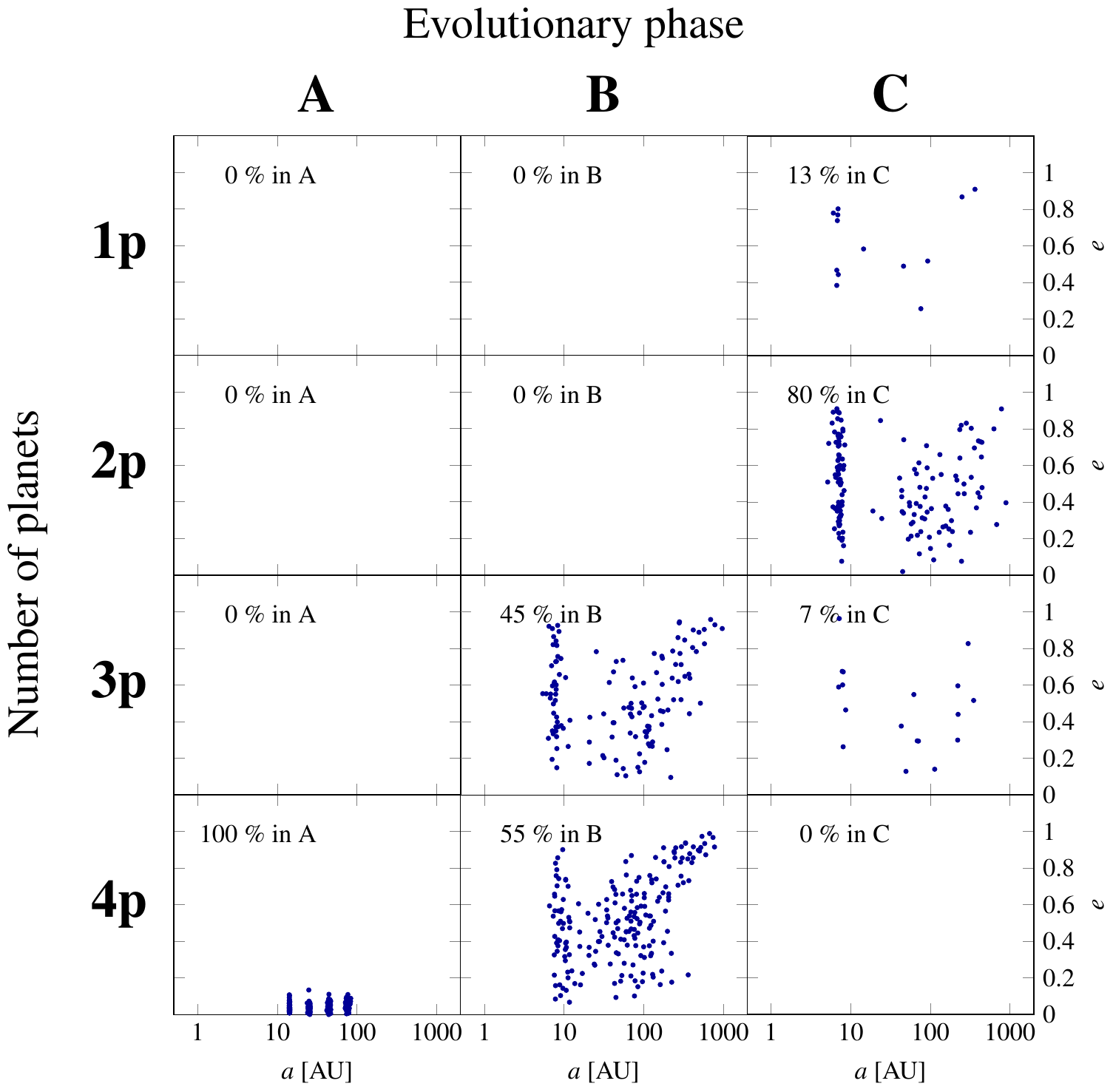}
\caption{Eccentricity and semi-major axes of the planets in each system of simulation~4f\_long at a randomly picked time for each system in each phase. From top to bottom we show the cases of 1-, 2-, 3-, and 4-planet systems and from left to right we show phase A, B, and C. All phase~A systems have low eccentricity and four planets. In phase~B, the systems have three or more planets. Typically, systems in phase~C have two planets, 
one at $a\sim 10$~AU and one at $30 \lesssim a \lesssim 1000$~AU. Both phase~B and C show generally higher eccentricities.}
\label{fig:ae}
\end{figure*}

\figref{fig:ae} shows how the eccentricity and semi-major axes of the planetary orbits change between phase~A, B, and C. In the figure we picked one random time in phase~A, B, and C  in each system in simulation~4f\_long, respectively.

Phase~A is characterised by low-eccentricity orbits, not deviating much from the initial set-up. Phase~B and C show an even distribution of eccentricities ($0 < e < 1$) and planets pile-up around $a\sim 10$~AU. All systems in phase B have three planets or more, while the most common number of planets in phase~C is two planets. In the two-planet case of phase~C, one planet typically has a semi-major axis of $a\sim 10$~AU and the other $a\sim 30-1000$~AU. The timescale of phase~A is what we referred to as the stability timescale, which is the time it takes before the first close encounter occurs. This timescale is strongly dependent on orbital separation \citep{2014prpl.conf..787D}. The timescale of phase~B is independent of initial orbit separation and we measure it to vary between $10^4$ to $10^7$~years, peaking at $10^6$~years. These timescales for the clearing of wide-orbit planets are consistent with previous studies \citep{2009ApJ...696.1600V}. This timescale is typically shorter than the duration of phase~A and the lifetime of the observed systems (see \tabref{tab:obs_planets}). Therefore statistically it is unlikely to observe a system in phase B unless the star is very young.

\subsection{Application to observed wide-orbit planetary systems}\label{sec:obs_systems}

\begin{table*}
\caption{Observed wide-orbit planets around stars.}             
\label{tab:obs_planets}      
\centering          
\begin{tabular}{lllcclcl} 
\toprule\midrule
System 	& Age & Spectral type & Planets & $r_{\mathrm{proj}}$ & $M_p$ & Predicted phase & References\\ 
& [Myr] & & & [AU] & [$M_{\mathrm{Jup}}$] \\
\midrule                   

1RXS J1609 & 5
& M0$\pm 1$V
 & b & $\sim$330
 & $8^{+3}_{-2}$
 & B/C 
 & 1, 2\\

AB Pic & $\sim$30 & K2V
& b 
& $\sim$260 & $13-14$ 
& C 
& 3\\

$\beta$ Pic & $12^{+8}_{-4}$
& A6V 
& b 
& $8-15$ 
& $9\pm 3$   
& A/C 
& 4, 5\\

Fomalhaut & $440$ 
& A3V 
& b$^{*}$ 
& $\sim 119$ & $\lesssim 3$
& B/C 
& 6, 7\\

GJ 504 & $160^{+350}_{-60}$ & G0V & b & 43.5 & $4.0^{+4.5}_{-1.0}$ & C 
& 8\\

HD 106906 & $13 \pm 2$ & F5V & b & $\sim 650$ & $11 \pm 2$ & B/C 
& 9, 10\\

HD 95086 
& $10 -17$  
& A8V 
& b & $56.4 \pm 0.7$ 
& $5 \pm 2$ 
 & A/B/C & 11, 12\\

HR~8799 & $30^{+20}_{-10}$ & A5V to F0V & b & 68 & $7^{+3}_{-2}$ & A 
& 13, 14, 15, 16\\
& & & c & 38 & $7^{+3}_{-2}$ & & 13, 14, 15\\
& & & d & 24 & $7^{+3}_{-2}$ & & 13, 14, 15\\
& & & e & 15 & $5 \pm 2$ & & 14, 15\\

$\kappa$ And & $30^{+20}_{-10}$ 
& B9IV 
& b 
& $55 \pm 2$ 
& $12.8^{+2.0}_{-1.0}$ 
& A/C 
& 17, 18\\

\bottomrule                
\end{tabular}
\tablefoot{1 -- \cite{2014ApJ...784...65B}, 2 -- \cite{2010ApJ...719..497L}, 3 -- \cite{2005A&A...438L..29C}, 4 -- \cite{2010Sci...329...57L}, 5 -- \cite{2001ApJ...562L..87Z}, 6 -- \cite{2012ApJ...754L..20M}, 7 -- \cite{2008Sci...322.1345K}, 8 -- \cite{2013ApJ...774...11K}, 9 -- \cite{2014ApJ...780L...4B}, 10 -- \cite{2012ApJ...746..154P}, 11 -- \cite{2013ApJ...772L..15R}, 12 -- \cite{2014A&A...565L...4G}, 13 -- \cite{2008Sci...322.1348M}, 14 -- \cite{2010Natur.468.1080M}, 15 -- \cite{2011ApJ...729..128C}, 16 -- \cite{1999AJ....118.2993G}, 17 -- \cite{2013ApJ...763L..32C}, 18 -- \cite{2011A&A...525A..71W}
\\
$^*$ -- Note news of Fomalhaut~b being dust cloud from collision with asteroid belt \citep{2011MNRAS.412.2137K,2014ApJ...786...70K}.
}
\end{table*}

In this section we predict the evolutionary phase of directly imaged, massive, wide-orbit planetary systems assuming their evolution is similar to that which we find for HR~8799 in simulation~4f\_long. \tabref{tab:obs_planets} gives some observed properties of the systems we include. 

\paragraph{1RXS J1609}

1RXS~J1609~b is found at a projected distance of 330~AU and is probably scattered to this distance as a protoplanetary disk of this size is unlikely. Its age is not sufficient to distinguish between phase~B or C, but if there are no other planets present we can predict phase~C.

\paragraph{AB Pic}

AB~Pic~b is also found at a very large distance from the star ($\sim 260$~AU). It is highly likely the planet has been scattered there and also taking the age of $\sim 30$~Myr into
account, we predict phase~C as phase~B generally lasts for shorter time.

\paragraph{$\beta$ Pic}

$\beta$~Pic~b is separated from $\beta$~Pic by $8-15$~AU. The fact that no outer planets have been found indicates either that the system is a one-planet system in phase~C or that the outer planets are not massive enough to be visible and the system is in phase~A or C. However, the low eccentricity of the orbit \citep[$e \sim 0.06$,][]{2014PNAS..11112661M} favours phase~A. 

\paragraph{Fomalhaut}

Fomalhaut~b is found at a projected distance of $\sim 119$~AU, which is far out and makes in situ formation less probable. The predicted eccentric orbit necessary to fit the observations would indicate phase~C considering the advanced age of the system ($\sim 440$~Myr). 

\paragraph{GJ 504}

The recently imaged planet around the solar-type star GJ~504 \citep{2008Sci...322.1345K} is found at a projected star-planet distance of $43.5$~AU. GJ~504~b could be an outer planet in a phase~C system (consider the age of $\sim 160$~Myr), which means that there might be another planet on a closer orbit. 

\paragraph{HD~106906}

HD~106906~b has an on-sky projected separation of $\sim 650$~AU, which makes it clear that the planet is scattered and the system is in phase~B or C as such a large protoplanetary disk is unlikely. It may be that there is also an inner planet in the HD~106906 system as the inner part cannot yet be resolved.

\paragraph{HD~95086}

HD~95086~b has a projected distance to the star of $56.4 \pm 0.7$~AU and could be an outer planet in a phase~B or C configuration, which then indicates possible inner companions. It could also be a phase~A system with an unresolved inner region because of the distance to the star \citep[$90.4\pm 3.4$,][]{2007A&A...474..653V}. The debris disk of HD~95086 is found to be similar to that of HR~8799 \citep{2015ApJ...799..146S}.

\paragraph{HR~8799}

The number of planets in HR~8799 and their low eccentricities make us classify the system as still being in phase~A.

\paragraph{$\kappa$ And}

The $\kappa$~And system seems like a higher mass version of the systems HD~95086 and GJ~504. The inner regions might hide a phase~A system whilst the outer regions might reveal an outer planet indicating phase~C. The system is old enough for us to predict phase~B to be unlikely.


\section{Summary and conclusions}\label{sec:conclusions}

The star HR~8799 is observed to be accompanied by four massive planets on wide orbits. Puzzlingly, the planetary system has been shown to be unstable in a timescale much shorter than its estimated age. A common explanation of its existence has been stabilisation by mean motion resonances \citep{2014MNRAS.440.3140G}.

We have simulated long-lived systems similar to HR~8799 without forcing them into resonance. The initial orbits of our planets are slightly more widely separated than circular orbits that intersect the observed star-planet on-sky separations. Slightly larger separations between the orbits result in a significantly longer stability timescale. The low eccentricity of the orbits can in some cases make a system have the observed architecture of HR~8799. We place the planets with random azimuth angle, assuming initial low eccentricities. 

When using the estimated planetary masses (5, 7, 7 and 7~$M_{\rm Jup}$) together with the estimated Earth-HR~8799 distance we rarely find a fitting system; in the best case, $\Delta _{\min} = 3.64-3.67,$ we find 5 of 100 fitting systems, and in the broader range of orbit separations we find 12 out of 900 fitting systems. We almost double the number of fitting systems by considering the $1\sigma$ longer distance to HR~8799 of 40.5~pc; the best case now considers orbit separation range $\Delta _{\min} = 3.74-3.79$ and yields 8 of 100 fitting systems: in total we find 22 of 900 fitting systems. It is easier to find a fitting system by considering slightly lower planet masses (4, 6, 6, and 6~$M_{\rm Jup}$) that are still comfortably within the planet mass estimate. That results in the best case ($\Delta _{\min} = 3.98 - 4.02$) 18 of 100 systems fit the observations and in the broader range of orbit separations we find 82 out of 900 systems fitting. The above described numbers are visualised in \figref{fig:frac8799surv}. Examples of a system that fits the observations can be seen for the nominal planet masses in \figrefthree{fig:sys94_sim4e}{fig:match}{fig:hist_stat} and for the slightly lower planet masses in \figreftwo{fig:tyre_245d}{fig:hist_245d}. 

We test our systems for mean motion resonances (see \secref{sec:MMR}) and find that most of them fall sporadically in and out of the 2:1 resonance between various planet pairs, however, we see no trend in long stability timescale in the systems with more resonant behaviour. This means that these resonances do not work in a stabilising way, unlike the resonances treated earlier in the literature for stability of the HR~8799 system. 

We conclude that the HR~8799 system does not necessarily have to be caught in a strong resonance. We simulated many systems that at some point in their evolution look like HR~8799 and we want to stress that our example system (system 94, simulation 4e) is not a lucky coincidence. We present a different solution than before as our fitting systems are stable for longer than the lifetime of HR~8799 and do not show signs that they are deep in resonance.

A bonus from our simulations is a prediction of the future evolution of the HR~8799 system. We see typical architectures in different evolutionary phases of the system. Before close encounters between the planets, the orbits have low eccentricity ($e < 0.2$). After the first close encounter has occurred, the first planet ejection follows within $\sim 10^6$ years. The close encounter and ejection change the architecture of the system and the eccentricities are now more evenly distributed ($0 < e < 1$). In $\sim80$\% of the cases, the systems eventually lose another planet and end up with two planets on eccentric orbits: one with a semimajor axis of $a_{\text{inner}}\sim 10$~AU and the other at $a_{\text{outer}}\sim 30-1000$~AU  (see \figref{fig:ae}). This configuration is comparably more similar to other massive, wide-orbit planets, such as AB~Pic~b or HD~106906~b, than the current HR~8799~bcde. 

Our HR~8799 analogues naturally evolve into architectures that resemble other wide-orbit planetary systems. Therefore we find it most plausible that the HR~8799 system is not of a different kind than other observed wide-orbit planetary systems, but simply in an earlier evolutionary phase. This phase seems to be relatively short for massive, wide-orbit planetary systems and it thus makes sense that only one system has been observed in this phase, that is HR~8799.


\begin{acknowledgements}
The authors thank the anonymous referees for their comments which improved the manuscript.
The authors thank Phil Uttley for expertise in statistics. YG acknowledges the support of a PhD studentship at the University of Amsterdam under the supervision of Selma E.\ de Mink. AJ and YG were partially funded by the European Research Council under ERC Starting Grant agreement 278675--PEBBLE2PLANET and by the Swedish Research Council (grant 2010–3710). MBD was supported by the Swedish Research Council (grant 2011-3991). RPC was supported by the Swedish Research Council (grants 2012-2254 and 2012-5807). Calculations presented in this paper were carried out at LUNARC using computer hardware funded in part by the Royal Fysiographic Society of Lund. AM was supported by the Swedish Research Council (grant 2011-3991) and KAW 2012.0150 from the Knut and Alice Wallenberg foundation. 
\end{acknowledgements}



\bibliographystyle{aa-package/bibtex/aa}
\bibliography{ref_planet2}

\Online

\begin{appendix} 

\section{Initial conditions of system 94, simulation 4e}

\hvFloat[%
nonFloat=true,%
capWidth=w,%
capPos=t,%
rotAngle=0,%
objectPos=c%
]{table*}{%
\begin{tabular}{lcccc}
\toprule \midrule
Planet & b & c & d & e \\
\midrule
Mass [$M_{\odot}$] & 0.004775549 & 0.006685769 & 0.006685769 & 0.006685769 \\ \\
$x$ [AU] & $-72.3056194919 $ & $-40.2180215763$ & $-13.0773477594 $ & $14.2619630629$ \\
$y$ [AU] & $-18.4404318999$ & $-17.5354438553$ & $21.3633582260$ & $0.0317906084$ \\
$z$ [AU] & $-0.4918207713$ & $0.1521487580$ & $-0.0232168947$ & $1.0418248166$ \\ \\
$v_x$ [AU / day] & $0.0006026600$ & $0.0012707737$ & $-0.0035903173$ & $-0.0000263986$ \\ 
$v_y$ [AU / day] & $-0.0023632584$ & $-0.0029135670$ & $-0.0021977681$ & $0.0055679781$ \\
$v_z$ [AU / day] & $0.0000077237$ & $0.0001137888$ & $0.0000052188$ & $0.0001914785$ \\
\bottomrule
\end{tabular}
}[]{%
Initial conditions of system 94, simulation 4e. This planetary system is so chaotic that these initial conditions are unlikely to reproduce our system 94, simulation 4e. A small change in timestep could alter the lifetime drastically.}{tab:ICs}

\end{appendix}

\end{document}